# Formation of Fe-6.5wt%Si High Silicon Steel by Double Glow Plasma Surface Metallurgy Technology

Zhong Xu[1, 2 a)], Jun Huang[3 a)], Hongyan Wu[4], Rui Chen[2], Chengyuan Zhang[2], Zaifeng Xu[1], Weixin Zhang[2], Lei Hu[2], Bin Zhang[2]

[1] Research Institute of Surface Engineering, Taiyuan University of Technology, Taiyuan 030024, China

[2] Apintec-Glow Technology Co.,Ltd, Suzhou 215100, China

[3] School of Materials Science and Engineering, Nanchang Hangkong University, Nanchang 330063, China

[4] Department of Material Physics, Nanjing University of Information Science and Technology, Nanjing 210044, China

Authors to whom correspondence should be addressed: xuzhong@tyut.edu.cn ,

huangjun@nchu.edu.cn

**Abstract:** High silicon steel with 6.5% silicon content is the best because of its excellent magnetic properties, such as high saturation magnetization, high resistivity, low iron loss and near zero magnetostriction. High silicon steel can greatly save energy, and reduce the weight and size of electrical appliances. This has a very important application prospect for energy and aerospace industry. The high brittleness of high silicon steel makes its production and processing very difficult.

For more than 30 years, many steel companies and research institutions around the world have adopted various technical means to study the industrialization of high silicon steel, but they have not been successful . JFE-NKK steel company in Japan has realized the small batch production of high silicon steel by using $SiCl_4$-CVD technology. However, due to the complex process, corrosion and pollution, high cost, its production scale is greatly limited. So far, large-scale production of high silicon steel is still a major challenge in the world. This paper will introduce the experimental results of successfully preparing high silicon steel by Double Glow Plasma Surface Metallurgy Technology. The process is simple and easy without any corrosion or pollution, which may provide a new way for the world to achieve large-scale production of high silicon steel. The large-scale production and wide application of high silicon steel is likely to change the pattern of the world's energy and electric power industry, save a lot of energy for mankind, and create huge economic benefits.

**Keywords:** Double Glow Plasma Surface Alloying; Fe-6.5wt%Si high silicon steel; homogeneous high silicon steel；Gradient high silicon steel; Xu-Tec Process。

## 1. Important significance of 6.5% silicon high silicon steel

Silicon steel is a steel material with iron as the matrix element and silicon as the main alloying element. Silicon steel is known as the artwork of steel materials and is an indispensable soft magnetic alloy in the power, electronics, and military industries, mainly used as iron cores for various motors, generators, and transformers. Its production process is relatively complex and regarded as the life of the enterprise, hence it is a highly confidential for domestic and foreign enterprises. The production technology and quality of silicon steel are regarded as one of the important indicators of a country's production and technological development level.

High silicon steel generally refers to Si-Fe alloys with a silicon content of 4.5 wt% to 6.7 wt%. High silicon steel with a silicon content of 6.5% is a premium product among high silicon steels. High silicon steel can significantly save energy, reduce material consumption, and enable many electrical appliances to achieve high efficiency and miniaturization.

In 1964, D. Brown[1] discovered that high silicon steel sheets with a silicon content of 6.5wt% Si exhibited excellent magnetic properties, such as low iron loss, near zero magnetostriction, and high saturation magnetization. However, high silicon steel with high silicon content has high hardness and brittleness, making it difficult to prepare through ordinary rolling methods. In order to avoid its brittleness problem, people use Technologies such as PVD, CVD, rapid condensation, hot rolling, powder rolling, salt bath siliconizing, etc to study and produce high silicon steel.

High silicon steel can significantly save energy, reduce material consumption, and enable many electrical appliances to achieve high efficiency and miniaturization. For example, in a transformer, an increase of 1.5% in energy efficiency implies a direct saving of $240 \times 10^9$ kW·h, which is equivalent to $12 billion/yr for electricity price of $0.05/kW·h[2]. Another example is that JFE Company in Japan uses high silicon steel containing 6.5wt% silicon instead of ordinary silicon steel containing 3 wt% silicon, reducing the weight of the iron core of an 8 kHz welding machine from 7.5 kg to 3 kg[3]. This is a very significant example of high silicon steel application. The slogan of China's aircraft manufacturing industry is "Fight to lighten one gram!" High silicon steel can greatly reduce the weight of aerospace electrical appliances and make them smaller, with very important application prospects.

For over 30 years, China's Beijing Iron and Steel Research Institute, Shanghai Baosteel, Capital Iron and Steel Company, Beijing University of Science and Technology, Taiyuan Iron and Steel Company, have invested a lot of funds (more than 10 billion yuan) and manpower in the research of high silicon steel, but have not achieved significant success.

## 2. Production lines for high silicon steel manufactured by NKK and JFE in Japan

In 1988, Japanese NKK Steel Company's Yoshikazu Takada and Masahiro Abe successfully produced high silicon steel containing 6.5% silicon in the laboratory using the $SiCl_4$-CVD method[4]. In 1993, NKK built a high silicon steel production line capable of manufacturing thicknesses of 0.1-0.5mm and widths of 400mm, with a monthly output of up to 100 tons[4]. In 1995, Japan successfully developed gradient high silicon steel JNHF Core based on the research of homogeneous high silicon steel JNEX Core.

The configuration of the line is shown in Fig. 1. The line consists of heating, CVD (siliconizing), diffusion, cooling, and coating zones. Cold-rolled, continuous, 3% silicon steel sheet, 600 mm wide and 0.05-0.3 mm thick, is used to produce coils of high-silicon content silicon steel sheet with an insulation coating.

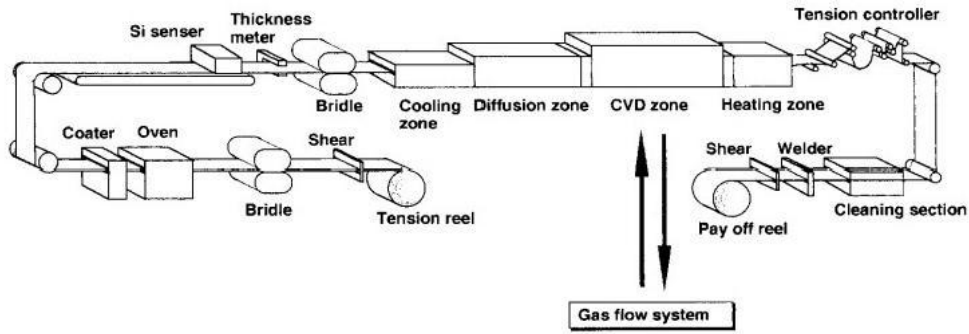

**Fig. 1.** Configuration of the continuous CVD siliconizing line [4].

However, there are some issues with using $SiCl_4$-CVD technology to manufacture high silicon steel, including:

(1) Performing at high temperatures (up to 1320 ℃) requires high equipment requirements and energy consumption.

(2) This production process is limited by the thickness of the sheet metal.

(3) $SiCl_4$ corrodes silicon steel sheets, causing corrosion pits on their surfaces, and the subsequent leveling process is cumbersome. The success yield of high silicon steel is relatively low.

(4) The process is complex, pollutes the environment, causes severe corrosion, has high costs, and is expensive.

The sales price of high silicon steel with 6.5% silicon content produced by JFE in Japan in the Chinese market is not only about 20 times that of ordinary high silicon steel, but also subject to very strict restrictions on its application.

## 3. Double Glow Plasma Surface Metallurgy Technology

Double Glow Plasma Surface Alloying Technology (Xu-Tec Process)[5-7], also known as Xu-Tec Process, was invented and is based on Plasma Nitriding Technology, which was developed by German scientist B. Berghuas in 1930. Xu-Tec process can form countless surface alloys with special physical and chemical properties on the surface of metal materials under low vacuum conditions, such as Cr alloy[8], W-Mo alloy[9], Ni-Cr alloy[10] and so on.

This plasma surface alloying technology has opened up a new field of "Plasma Surface Metallurgy, it is an interdisciplinary subject combining traditional materials science with Applied Physics. Xu-Tec process has been successfully applied to steel plate, sawing tools, flange, colloid mill and other industrial products. In recent years, in terms of X material refining, tantalum alloy has been formed on the surface of its core components by using this technology, thus increasing its service life by more than 20 times, solving a major engineering problem that has not been solved in China for more than 30 years.

Fig. 2 displays the Schematic Diagram of Xu-Tec siliconizing setup. Three electrodes are set in a vacuum vessel: anode（green）, work-piece cathode（blue）and two source cathode（purple）which are made by silicon materials. Two adjustable voltage DC power supplies (0–1200 V) are set between the anode and the work-piece and between the anode and the two source cathodes, respectively. Both the work-piece and the source are in negative potential. When the chamber is vacuumed, a certain amount of argon gas is filled. Then, with the increase in voltages of the power supplies, glow discharges will be generated between the anode and the work-piece and between the anode and the source at the same time. The positive ions of argon generated by glow discharge bombard the source cathode, causing the silicon to be sputtered and subsequently transported and adsorbed onto the surface of the sample. At the same time, the positive ions of argon bombard the sample cathode, heating it up to high temperature and causing the silicon adsorbed onto the surface of the sample to diffuse inward, forming a layer rich in silicon[11, 12]. In the figure, the left side is the experimental device, and the right side is the results obtained after siliconizing treatment: the upper curve is the distribution curve of silicon content of homogeneous high silicon steel in its section, and the lower curve is the distribution curve of silicon content of gradient high silicon steel in its section.

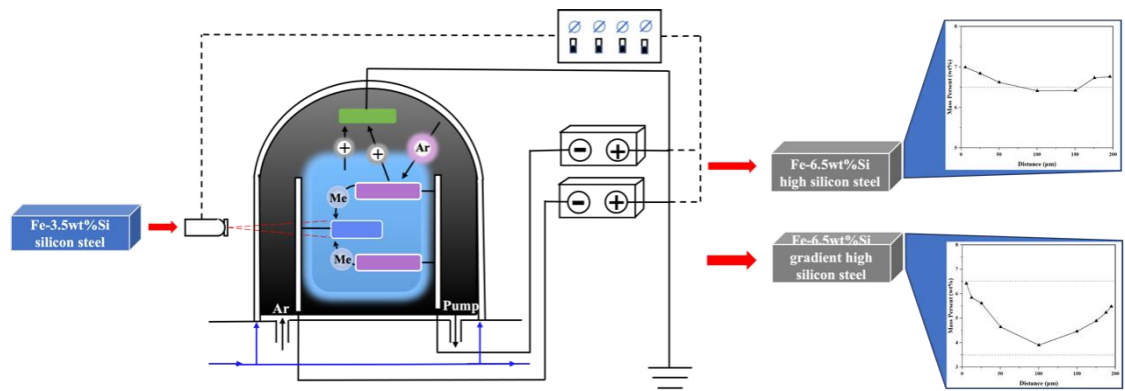

**Fig. 2.** Schematic Diagram of Xu-Tec siliconizing setup unit.

On this purpose, the experimental sample is made of Fe-3.6wt%Si silicon steel sheet with 0.20 mm and size of 60 mm×45 mm. The chemical composition of the 3.6 wt% silicon steel is listed in Table 1. The sample surface was mechanically polished by SiC abrasive paper up to 2000 grit, and then ultrasonically cleaned in alcohol for 15 min. A pure silicon plate (size 100 mm × 80 mm × 6 mm) was used as the sputtering target, called source electrode, and then Silicon elements were diffused into the 3.6wt% silicon steel, called work-piece electrode, by Xu-Tec process. Field emission scanning electron microscopy and energy spectrometer were used to characterize the surface and cross-sectional morphology of the coating, the corresponding elemental composition was also detected. The high silicon steels produced by JFE company in Japan, JNEX 900 and JNHF 600 were also detected as a comparison sample.

**Table 1** Chemical composition of the silicon steel (3.6 wt% Si) sheet.

| Sample | Composition (wt%) | |
|---|---|---|
| | Si | Fe |
| Silicon Steel | 3.60 | 97.40 |

Experimental materials and main process parameters:

1. Work-piece material: Silicon steel sheet containing 3.6wt% silicon, with dimensions of 70mm X 70mm X 0.2mm.

2. Source material: 6mm thick monocrystalline silicon

3. Argon gas pressure: 15-35 Pa;

4. Source voltage：900-1100 V;

5. Working-piece voltage：300-600 V;

6. The distance between the working-piece and the source: 16-35 mm;

7. Operation alloying temperature：1000-1200 ℃;

8. Holding time：1-6 h.

## 4. Experimental results

Fig. 3 shows the cross-sectional microstructure and silicon content of Fe-6.5wt%Si high silicon steel. Fig. 3a and 3b were the cross-sectional morphology and the silicon content distribution curve of Xu-Tec high silicon steel, while Fig. 3c and 3d were those of JNEX 900. It is obvious that both Xu-Tec high silicon steel and JNEX 900 have an average silicon content in their cross-sections that is more than 6.5 wt%. The surface silicon content of Xu-Tec high silicon steel is 6.99 wt% on one surface and 6.76 wt% on the other, with a central content of 6.41 wt%. However, it should be noted that thickness of Xu-Tec high silicon steel is twice of JNEX 900, which means that Xu-Tec has much better element diffusion efficiency.

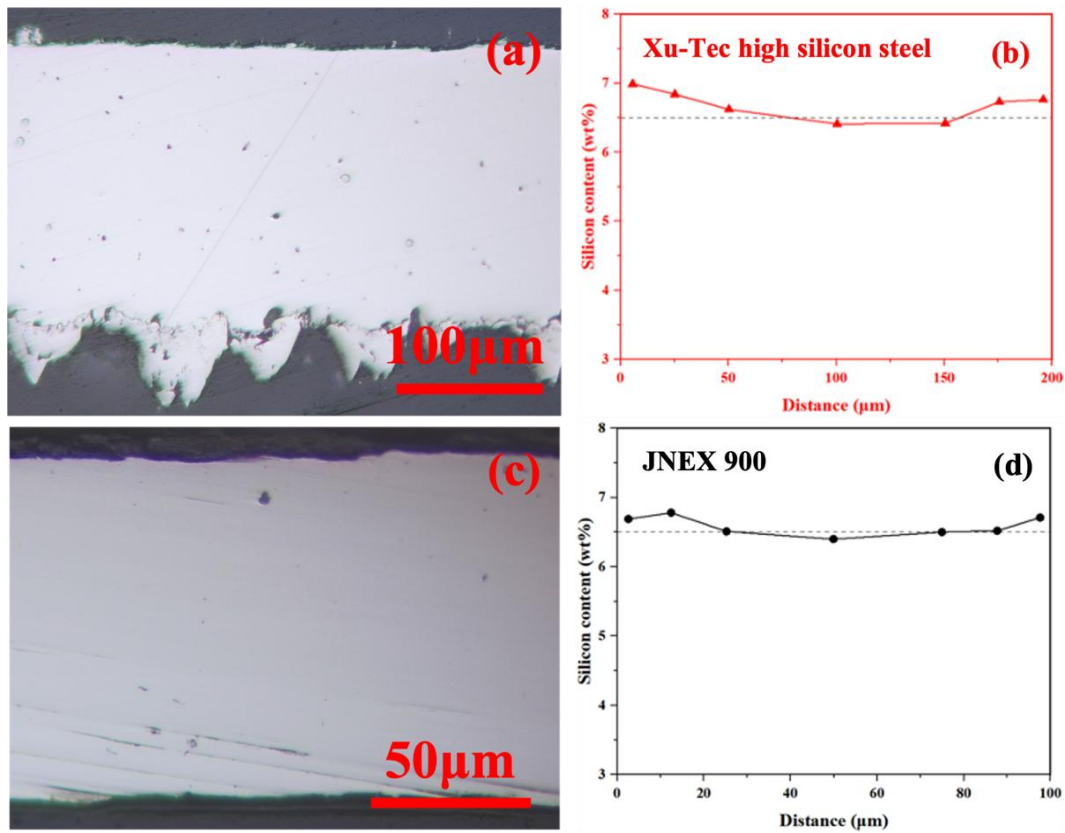

**Fig. 3.** Microstructure and silicon distribution of cross-section for Xu-Tec high silicon steel (a and b) and JNEX 900 (e and d).

According to the above experimental results, 5 times repeated experiments were carried out, and their silicon content in cross-section were shown in Fig. 4. The results show that there is good repeatability. Fig. 4b present the average silicon content of Xu-Tec high silicon steels in cross-section. There is no significant difference in the average silicon content among 1/4 depth, 1/2 depth and the core, meaning that the silicon content in Xu-Tec high silicon steel is homogeneous. There is also no significant difference in the average silicon content between the surface and 1/4 depth. According to one-way analysis of variance (ANOVA) was performed to determine the statistical significance of the data. Differences were considered significant at $P < 0.05$, and highly significant at $P < 0.01$.

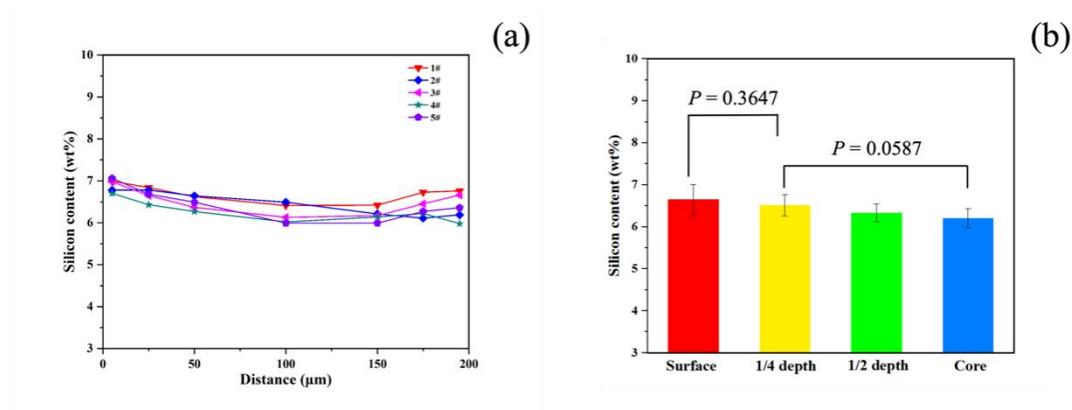

**Fig. 4**. Silicon content of Xu-Tec high silicon steels in cross-section.

Fig. 5 shows microstructure and silicon distribution of Fe-6.5wt%Si gradient high silicon steel. The cross-sectional morphology and the silicon content distribution curve of Xu-Tec gradient high silicon steel were shown in Fig. 5a and 5b, while those of JNHF 600 were illustrated in Fig. 5c and 5d. A high silicon steel with gradient silicon content in cross-section was also fabricated by Xu-Tec process. The surface silicon content of Xu-Tec gradient high silicon steel is 6.42 wt% on one surface and 5.48 wt% on the other, with a central content of 3.90 wt%.

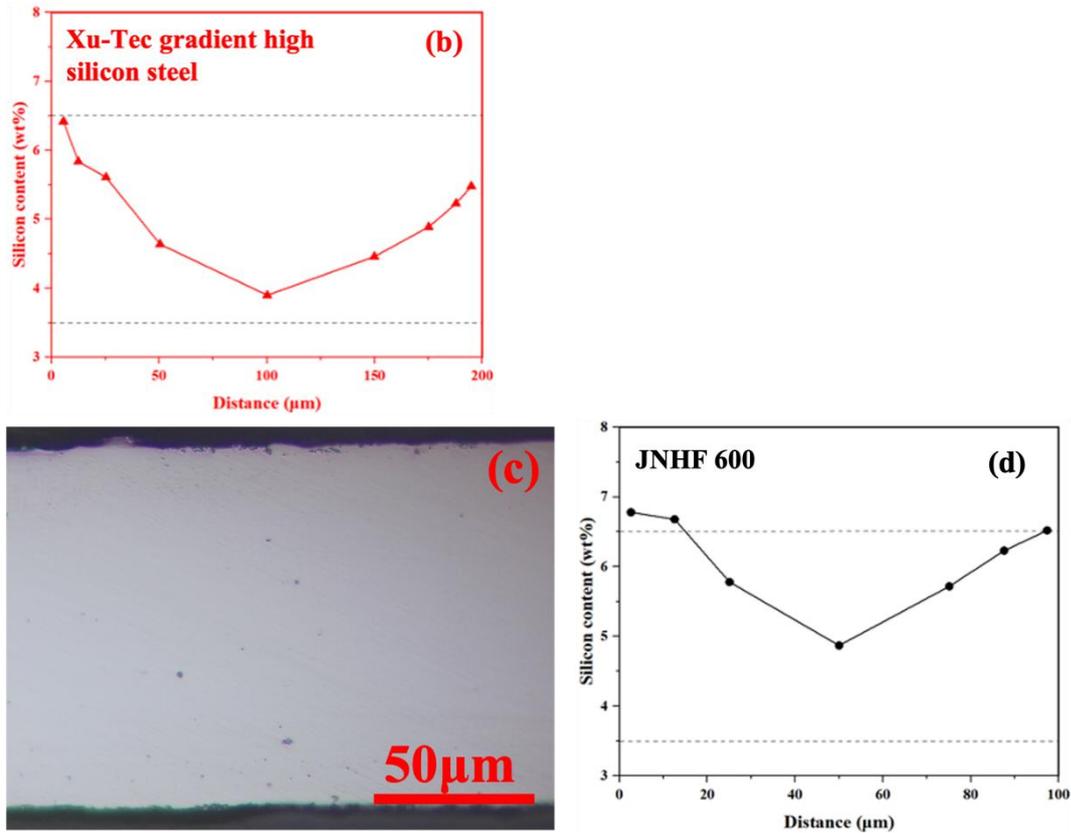

**Fig. 5.** Microstructure and silicon distribution of cross section for gradient high silicon steel Xu-Tec (a and b) and JNHF 600 (c and d).

The influence of alloying time on silicon content and iron loss were studied, and their relationship was illustrated in Fig. 6. The experimental results show that the homogeneous high silicon steel containing 6.5% silicon can be obtained by siliconizing with the holding time of 6 hours, and its iron loss at 50Hz is 0.68. The rest high silicon steel obtained under 6-hour holding time is gradient high silicon steel with different silicon content distribution, and its iron loss at 50 Hz is less than 1.0 W/kg, which is much lower than that of the untreated 3.6 wt% silicon steel. The iron loss of raw material containing 3.6% silicon steel at 50Hz is

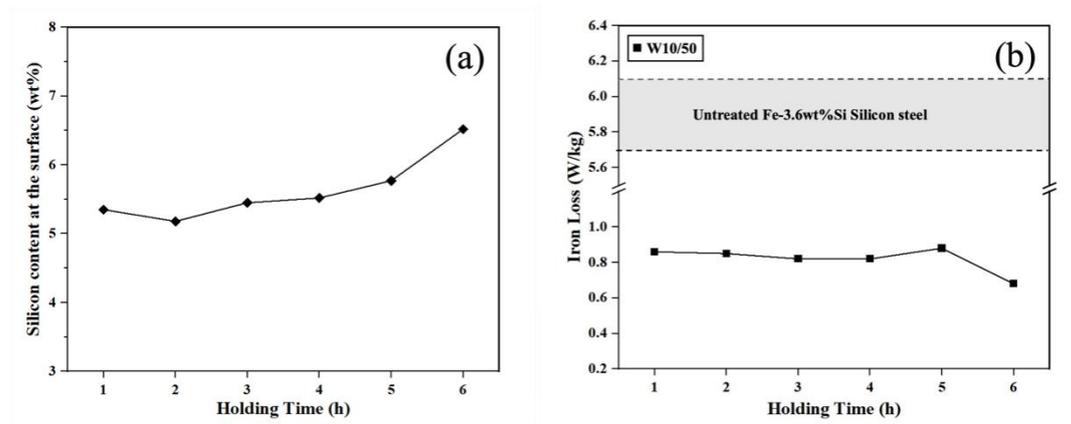

**Fig. 6.** The curves of holding time vs. silicon content and iron loss.

The Xu-Tec siliconizing is relatively simple, easy to implement, without any corrosion and pollution, and may provide a new way to achieve large-scale production of high-silicon steel in the world.

## 5. Conclusion and Prospect

1. Research results show that the Xu-Tec process is feasible to prepare high silicon steel.

2. Feasibility of using mono-crystalline silicon as a sputtering target to provide silicon element.

3. It is proved that low silicon steel is feasible to produce high silicon steel as raw material of high silicon steel.

4. It is proved that the low silicon steel with thickness of 200 μm can directly form uniform high silicon steel.

5. Except for uniform high silicon steel, the rest belong to the range of gradient high silicon steel. It is easy to prepare gradient high silicon steel.

6. Xu-Tec siliconizing does not have high temperature corrosion and environmental pollution in the production process. Under the condition of ion bombardment, the diffusion rate of silicon is fast.

7. Experimental study provides an experimental basis for the application research and industrialization of high silicon steel in the future.

8. Xu-Tec technology provides a new way for the development and industrialization of high silicon steel, and its prospect is very bright.

Our research team originally planned to realize the industrialization of high silicon steel through the following four stages: Basic feasibility study - Application Study - Pilot stage - Large-scale production stage. We have completed the first stage of experimental research, as we have shown the result in this article. We hope to generate more interest and support.

## Acknowledgement

The authors is very grateful to young entrepreneurs Mr. Bin Zhang for investing our research on high silicon steel.